\documentclass[useAMS,usenatbib]{mn2e} 
\usepackage{graphicx}
\usepackage{url} 

\newcommand{\MSun}{\mbox{${\rm M}_\odot$}}
\newcommand{\Msun}{\mbox{${\rm M}_\odot$}}

\def\lteq{\ {\raise-.5ex\hbox{$\buildrel<\over-$}}\ }
\def\apgt{\ {\raise-.5ex\hbox{$\buildrel>\over\sim$}}\ }
\def\aplt{\ {\raise-.5ex\hbox{$\buildrel<\over\sim$}}\ }
\def\lt{\ {\raise-.5ex\hbox{$\buildrel>$}}\ }
\def\gt{\ {\raise-.5ex\hbox{$\buildrel<$}}\ }
\def\eqgt{\ {\raise-.5ex\hbox{$\buildrel>\over-$}}\ }
\def\eqlt{\ {\raise-.5ex\hbox{$\buildrel<\over-$}}\ }

\newfont{\Giga}{cmssbx10 scaled 5200}
\newfont{\giga}{cmssbx10 scaled 4300}
\newfont{\Mega}{cmssbx10 scaled 3200}
\newfont{\mega}{cmssbx10 scaled 2500}
\newfont{\Kilo}{cmssbx10 scaled 2000}
\newfont{\kilo}{cmssbx10 scaled 1600}
\newfont{\Deca}{cmssbx10 scaled 1450}
\newfont{\deca}{cmssbx10 scaled 1200}
\newfont{\Dezi}{cmssbx10 scaled 1100}
\newfont{\dezi}{cmssbx10 scaled 1050}
\newfont{\iGiga}{cmssi10 scaled 6200}
\newfont{\igiga}{cmssi10 scaled 4300}
\newfont{\iMega}{cmssi10 scaled 3200}
\newfont{\imega}{cmssi10 scaled 2500}
\newfont{\iKilo}{cmssi10 scaled 2000}
\newfont{\ikilo}{cmssi10 scaled 1500}
\newfont{\mathGiga}{cmsy10 scaled 6200}
\newfont{\mathgiga}{cmsy10 scaled 4300}
\newfont{\mathMega}{cmsy10 scaled 3200}
\newfont{\mathmega}{cmsy10 scaled 2500}
\newfont{\mathKilo}{cmsy10 scaled 2000}
\newfont{\mathkilo}{cmsy10 scaled 1500}
\newfont{\mathDeca}{cmsy10 scaled 1450}
\newfont{\mathdeca}{cmsy10 scaled 1200}

\def\aap{\ {A\&A}\ }

\def\aj{\ {AJ}\ }

\def\apj{\ {ApJ}\ }
\def\apjl{\ {ApJL}\ }

\def\araa{\ {ARA\&A}\ }

\def\icarus{\ {Icarus}\ }

\def\mnras{\ {MNRAS}\ }

\def\apgt{\ {\raise-.5ex\hbox{$\buildrel>\over\sim$}}\ }
\def\aplt{\ {\raise-.5ex\hbox{$\buildrel<\over\sim$}}\ }
\def\lteq{\ {\raise-.5ex\hbox{$\buildrel<\over-$}}\ }

\title[]{Stellar disk destruction by dynamical interactions in the Orion Trapezium star cluster}

\author[Simon Portegies Zwart]
{Simon~F.~Portegies~Zwart$^{1}$\\
$^{1}$Leiden Observatory, Leiden University, Leiden, The Netherlands\\
}
\begin{document}

\date{}
\maketitle

\begin{abstract} 
We compare the observed size distribution of circum stellar disks in
the Orion Trapezium cluster with the results of $N$-body simulations
in which we incorporated an heuristic prescription for the evolution
of these disks. In our simulations, the sizes of stellar disks are
affected by close encounters with other stars (with disks).  We find
that the observed distribution of disk sizes in the Orion Trapezium
cluster is excellently reproduced by truncation due to dynamical
encounters alone.  The observed distribution appears to be a sensitive
measure of the past dynamical history of the cluster, and therewith on
the conditions of the cluster at birth.  The best comparison between
the observed disk size distribution and the simulated distribution is
realized with a cluster of $N = 2500\pm500$ stars with a half-mass
radius of about 0.5\,pc in virial equilibrium (with a virial ratio of
$Q = 0.5$, or somewhat colder $Q \simeq 0.3$), and with a density
structure according to a fractal dimension of $F \simeq 1.6$.
Simulations with these parameters reproduce the observed distribution
of circum stellar disks in about 0.2--0.5\,Myr.

\end{abstract}


\section{Introduction}

The Trapezium cluster in the Orion nebula
\cite{Huygens1656,Huygens1899} (later named M\,42, NGC\,1976) is one
of the closest 412\,pc \citep{2009ApJ...700..137R} young $\sim
0.3$\,Myr \cite[85\% of the stars $\aplt 1$\,Myr][but see also
  \cite{1997AJ....113.1733H,1998ApJ...492..540H}]{1994ApJ...421..517P}
star forming regions, composed of about $10^3$ stars within a radius
of $\sim 3$\,pc \citep{1999AJ....117..354D}. Even though the cluster
is nearby and about to emerge from its parental molecular cloud
\citep{2013A&A...549A.114L}, its age, the number of members and the
origin of its spatial and kinematic structure remain uncertain.  Being
one of the closest relatively massive young stellar systems it forms a
key to understand cluster formation and early evolution.

The close proximity of the Trapezium cluster allows detailed
observations of circum-stellar disk sizes using HST/WFPC2
\citep{2005A&A...441..195V}.  This size distribution is well
characterized by a power-law \citep{2005A&A...441..195V}, but the
origin of this distribution remains uncertain.  Dynamical interactions
in young clusters have been demonstrated to be important for the sizes
of circum stellar disks \citep{2015A&A...577A.115V}, and the majority
of protoplanetary disks are likely to be truncated by close stellar
encounters.  It is however, not clear whether in the Trapezium this
process can still be recognized in the observed distribution of circum
stellar disks.  \cite{2005A&A...441..195V} argue that: {\em albeit the
  young age of the Trapezium, and given that disk destruction is well
  underway, it is perhaps too late to tell if the present day disk
  size distribution is primordial or if it is a consequence of the
  massive star formation environment.}

Here we show that the size distribution of the observed circum stellar
disks is consistent with the disk size distribution that result from
close encounters in young star clusters born with complex structure.
We subsequently use the observed distribution of disk sizes to
reconstruct the history of the dynamical and kinematic and structure
of the cluster.

\section{Methods}

We use the Astronomical Multipurpose Software Environment
\citep[AMUSE][]{2009NewA...14..369P,2013CoPhC.183..456P,2013A&A...557A..84P}
to carry out simulations.  AMUSE allows us to generate initial
conditions, combine a wide range of gravitational $N$-body packages
and stellar evolution modules together with other physical models, and
process the data.

The application script is written in python, even though the
scientific production codes are written in compiled languages. In this
way, the generation of initial conditions and data processing is
mostly done at the relatively slow script-level, whereas the most
demanding tasks are carried out with optimized code for high
performance.  The overhead introduced by opting for a scripting
language for the data management is negligible.

Our production script starts by generating initial conditions for the
young star cluster.  The gravitational calculations are solved using
the 4th-order Hermite $N$-body code {\tt ph4} (Steve McMillan, private
communication), with an time-step parameter $\eta = 0.01$ and a
softening of 100\,AU.

During the integration of the equations of motion, we check for close
approaches. When two stars happen to approach each other in a
pre-determined encounter radius (initially 0.02\,pc) we interrupt the
$N$-body integrator after synchronizing the system to subsequently
resolve the encounter.

\subsection{The effect of encounters on disk size}\label{Sect:DiskRadius}

The effect of the two-body encounter on the disks of both stars is
solved semi-analytically.  Once a two-body encounter is detected we
calculate the pericenter distance, $q$, by solving Kepler's equation
\cite[using the kepler-module from the {\tt Starlab}
  package,][]{2001MNRAS.321..199P}.  Note that the closest approach
may be well within the adopted softening radius of 100\,AU.  The new
disk radius for a star with mass $m$ is calculated using
\citep[][which was calibrated for parabolic co-panar prograde
  encounters]{2014A&A...565A.130B}:
\begin{equation}
  r^\prime_{\rm disk} = 0.28 q \left( {m \over M} \right)^{0.32}.
\end{equation}
Here $M$ is the mass of the other star.  This equation is also applied
for calculating the new disk radius of the encountering star.  These
new radii are adopted only if they are smaller than the pre-encounter
disk radii. This procedure does not affect the dynamics of the system
in the sense that the stars are not moved, although their total mass
(star plus disk) is affected before the actual pericenter passage.

In order to reduce the number of disk truncations at runtime, and
therewith the number of interrupts (and synchronizations) in the
$N$-body integration, the new encounter distance for both stars is
reset to half the pericenter distance.  This prevents two stars from
being detected at every integration time step while approaching
pericenter, which would cause the disk to be affected repeatedly
during a single encounter.  This procedure therefore limits the number
of encounters to the most destructive one at pericenter.

\subsection{The effect of encounters on disk mass}

The truncated disks of the encountering stars lose mass.  We estimate
the amount of mass lost from each disk using
\begin{equation}
    \mathrm{d}m = m_{\rm disk} 
          {r_{\rm disk}^{1/2}-r^{\prime 1/2}_{\rm disk} \over r_{\rm disk}^{1/2}}.
\end{equation}
Both encountering stars may accrete some of the material lost from the
other star's disk, which we calculate with
\begin{equation}
  \mathrm{d}m_{\rm acc} = \mathrm{d}m f {m \over M+m}.
\label{Eq:dmdisk}\end{equation}
Here $f \leq 1$ is a mass transfer efficiency factor.  Both equations
are applied symmetrically in the two-body encounter, and as a
consequence both stars lose some mass and gain some of what the other
has lost. 

After every 0.1\,Myr we synchronize the gravity solver, check for
energy conservation, and dump a snapshot to file for later analysis.
The energy of the $N$-body integrator is preserved better than
$1/10^8$, which is sufficient to warrant a reliable result
\citep{2041-8205-785-1-L3}.


\section{Results}

\subsection{Initial conditions}

Each calculation starts by generating a realization for the $N$-body
model: stellar masses, positions and velocities.  Each star is
subsequently provided with a disk of 10\,\% of the stellar mass and
with an outer radius of $r_{\rm disk} = 400$\,AU.  This corresponds to
the maximal disk radius observed in the Trapezium cluster
\citep{2005A&A...441..195V}. It seems a bit small compared to proto
stellar disk sizes \citep{2009ApJ...700.1502A,2010ApJ...723.1241A},
but for our calculations we only want to know if they are truncated
below the maximum observed radius.

The choice of disk mass is somewhat arbitrary, but not inconsistent
with observed masses of young proto-planetary disks
\citep{2011ARA&A..49...67W}. The mass of the initial disk is added to
the stellar mass for the $N$-body integration.  The change in disk
mass due to encounters is self consistently taken into account during
the integration.

Stellar masses are selected randomly from a broken power-law
\citep{2001MNRAS.322..231K} between 0.01\,\Msun, and 100\,\MSun\, (The
mean mass of this mass function $\langle m \rangle \simeq
0.396$\,\MSun).  The positions of the stars are selected from a
fractal distribution \citep{2004A&A...413..929G} with a fractal
dimension of $F=1.2$, 1.6 and $F=2.0$, but additional simulations were
performed using the \cite{1911MNRAS..71..460P} distribution. The
velocities of the stars ware initially scaled such that the cluster
has virial ratio $Q=0.1$, $0.3$, $0.5$, $0.7$ and $Q=1.0$. A value of
$Q<0.5$ results in rather cold initial conditions, $Q \equiv 0.5$ puts
the cluster in virial equilibrium, and higher values are suitable for
super-virial clusters.  We varied the number of stars (from 1000 to
3500 in steps of 500) and the initial characteristic cluster radius
(from 0.125\,pc in steps of 2 to 1\,pc).

\begin{figure}
\begin{center}
\includegraphics[width=0.5\textwidth]{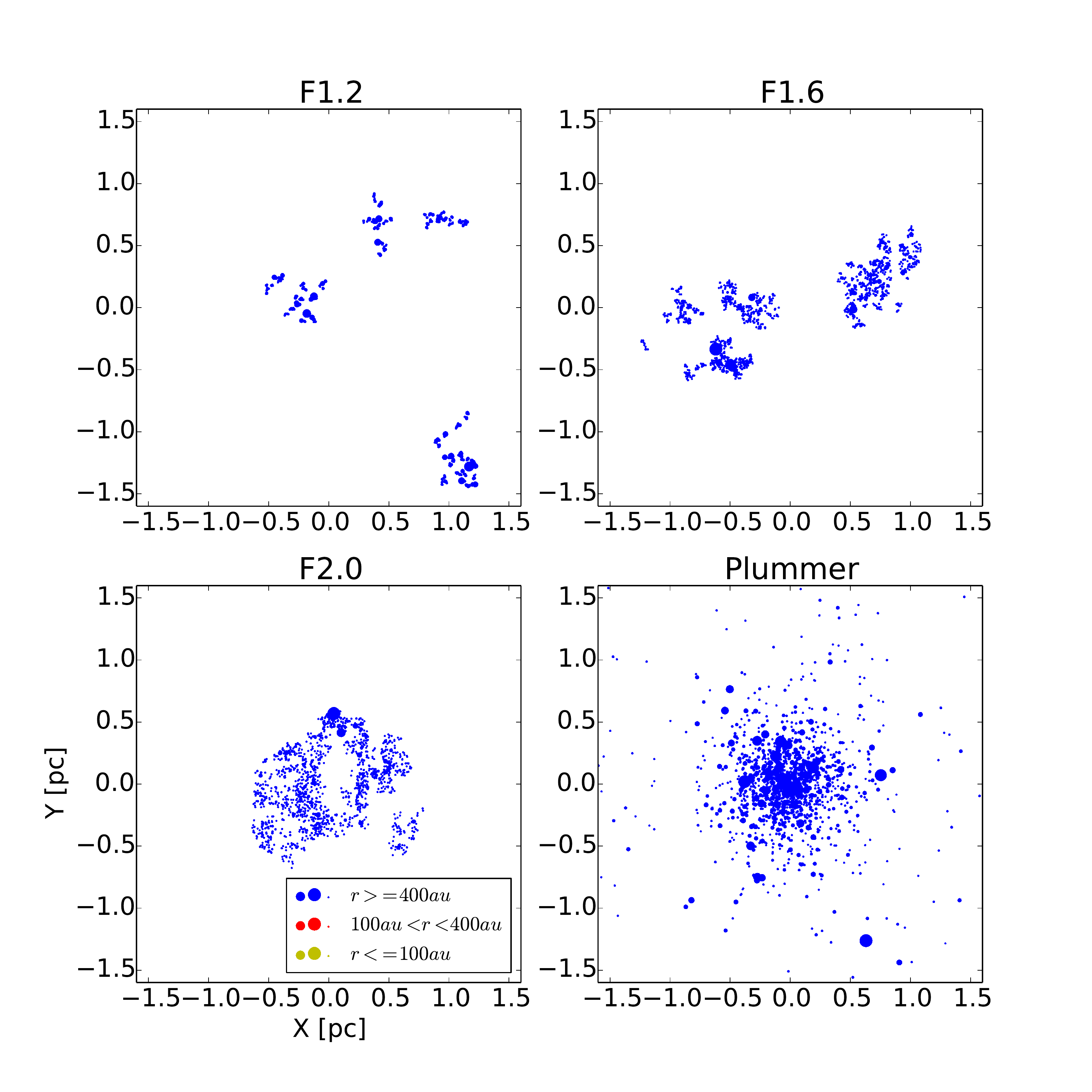}
\end{center}
\caption{Initial conditions for 4 clusters, each composed of $N=1500$
  stars initially in virial equilibrium (Q=0.5) and distributed with a
  characteristic radius of 0.5\,pc. From the top left to the bottom
  right give a fractal distribution with F=1.2, F=1.6 (top right),
  F=2.0 (bottom left) and a Plummer sphere (bottom right)
\label{Fig:InitialConditions}
}
\end{figure}

\begin{figure}
\begin{center}
\includegraphics[width=0.5\textwidth]{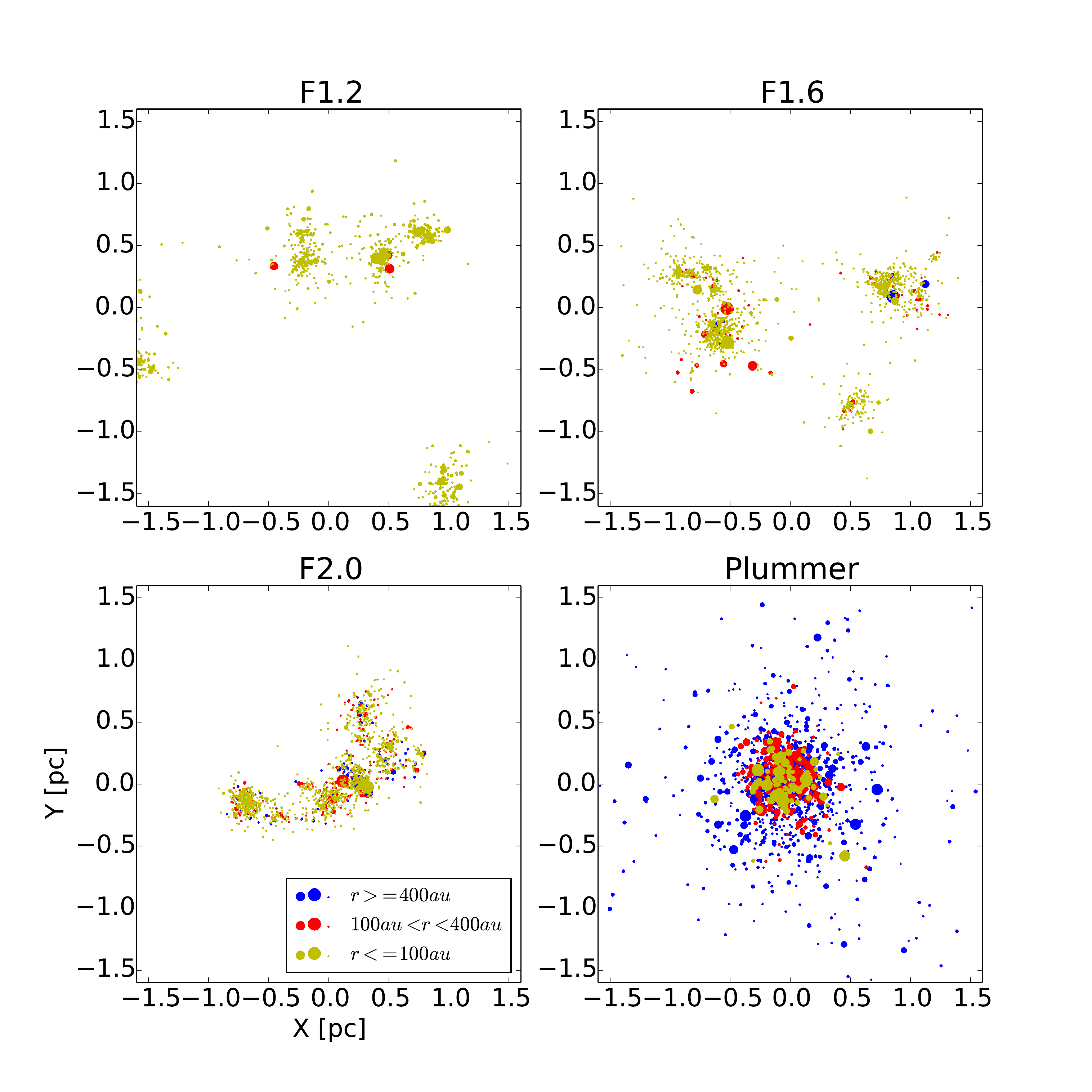}
\end{center}
\caption{Presentation of four clusters from the initial conditions
  which we presented in Fig.\,\ref{Fig:InitialConditions}, but evolved
  to 0.3\,Myr.The various colors indicate the limiting radii of their
  disks (see bottom left for the legend).
\label{Fig:FinalConditions}
}
\end{figure}

We present an impression of the various initial conditions in
Fig.\,\ref{Fig:InitialConditions}, and the consequence of the
dynamical evolution after 0.3\,Myr in
Fig.\,\ref{Fig:FinalConditions}.

Every calculation was performed four times with a different random
seed to generate the initial realization.  In addition, the models
that compared best to the observations have $R=0.5$\,pc, $Q=0.3$ (and
$Q=0.5$), $F=1.6$ were performed 12 times for each value of $N$, and
with an output time-resolution of 0.02\,Myr.

\subsection{Disk size distribution}

All calculations were stopped at an age of 1\,Myr.  During this period
the gravitational dynamics of the stars is resolved numerically using
Newtons law of motion. Close encounters result in the truncation of
the circum stellar disks. Due to the absence of any other disk
destruction mechanism all the evolution in the disks is the result of
the dynamical encounters, and our simple disk-destruction prescription
(see \S\,\ref{Sect:DiskRadius}).

\begin{figure}
\begin{center}
\includegraphics[width=0.5\textwidth]{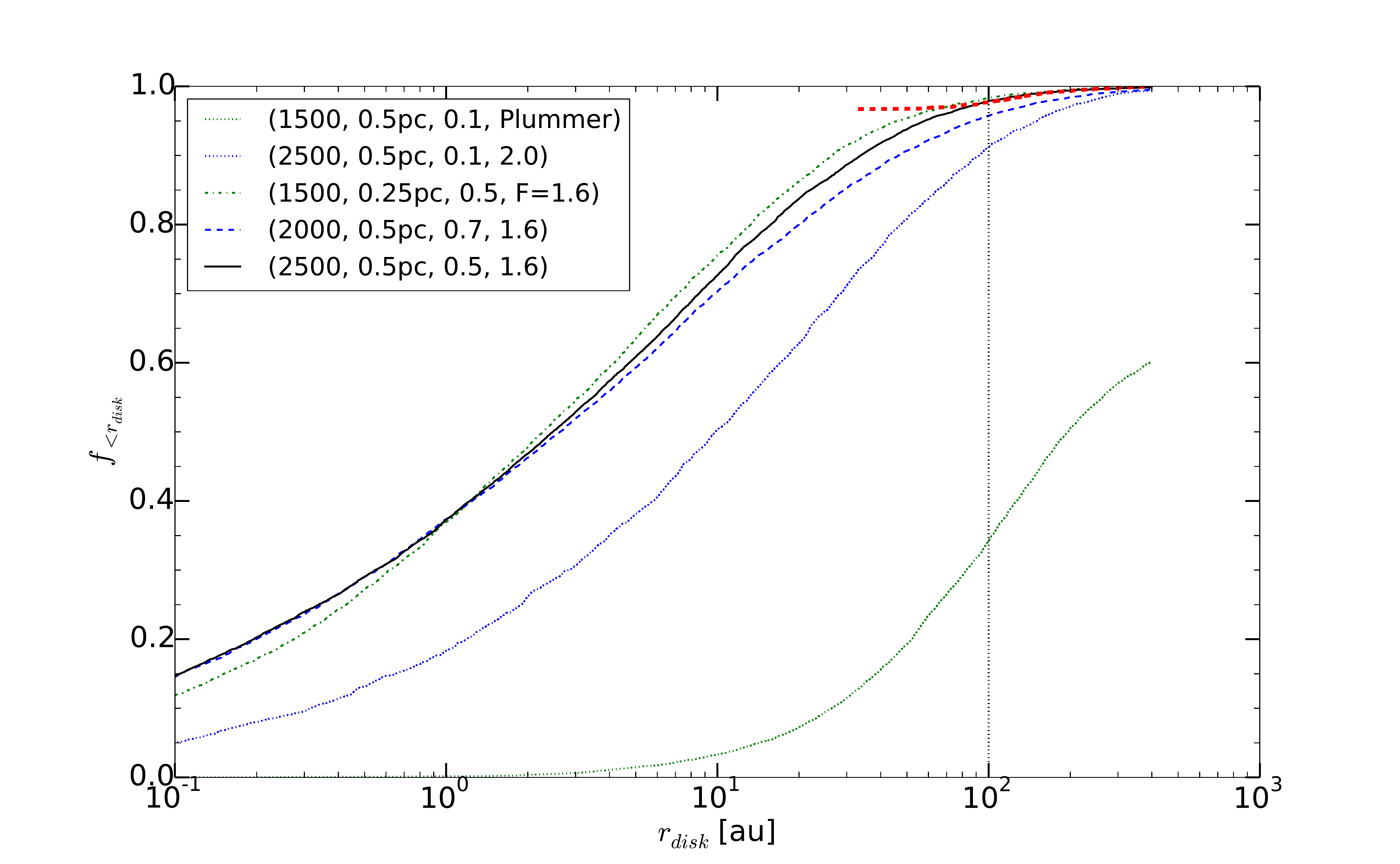}
\end{center}
\caption{Cumulative radius distribution of circum-stellar disks from
  several simulations (see top left for the legend with number of
  stars $N$, radius $R$ in parsec, virial temperature $Q$ and fractal
  dimension $F$).  The completeness limit in the observations at
  100\,AU is indicated with the vertical black dotted curve.  In the
  simulations we are not plagued by observational election effects.
  The red dotted curve gives the observed disk distribution, scaled to
  the model with the most comparable disk distribution (solid black
  curve) and with the vertical offset for disk radii $\geq 100$\,AU.
\label{Fig:SimulatedDisDistribution}
}
\end{figure}

In fig.\,\ref{Fig:SimulatedDisDistribution} we present the size
distribution of the disks from several of the simulations.  To
illustrate the wide range in disk distributions depending on the
initial conditions of the simulations, we show one excellent
comparison, and several less satisfactory cases.

The observed distribution is limited by the telescopes resolution
\citep{2005A&A...441..195V}.  The pixel size in these observations is
45\,AU, and they are able to resolve disks at 1.5 pixels (or 67.5\,AU),
although they argue that their sample is complete down to a minimum
radius of 100 to 150\,AU.  From the 162 proplyds 95 are larger than
100\,AU \citep{2005A&A...441..195V}.  For the analysis we 
compare observed disk sizes with the simulated distribution of disks
with a radius of at least 100\,AU.

\begin{figure}
\begin{center}
\includegraphics[width=0.5\textwidth]{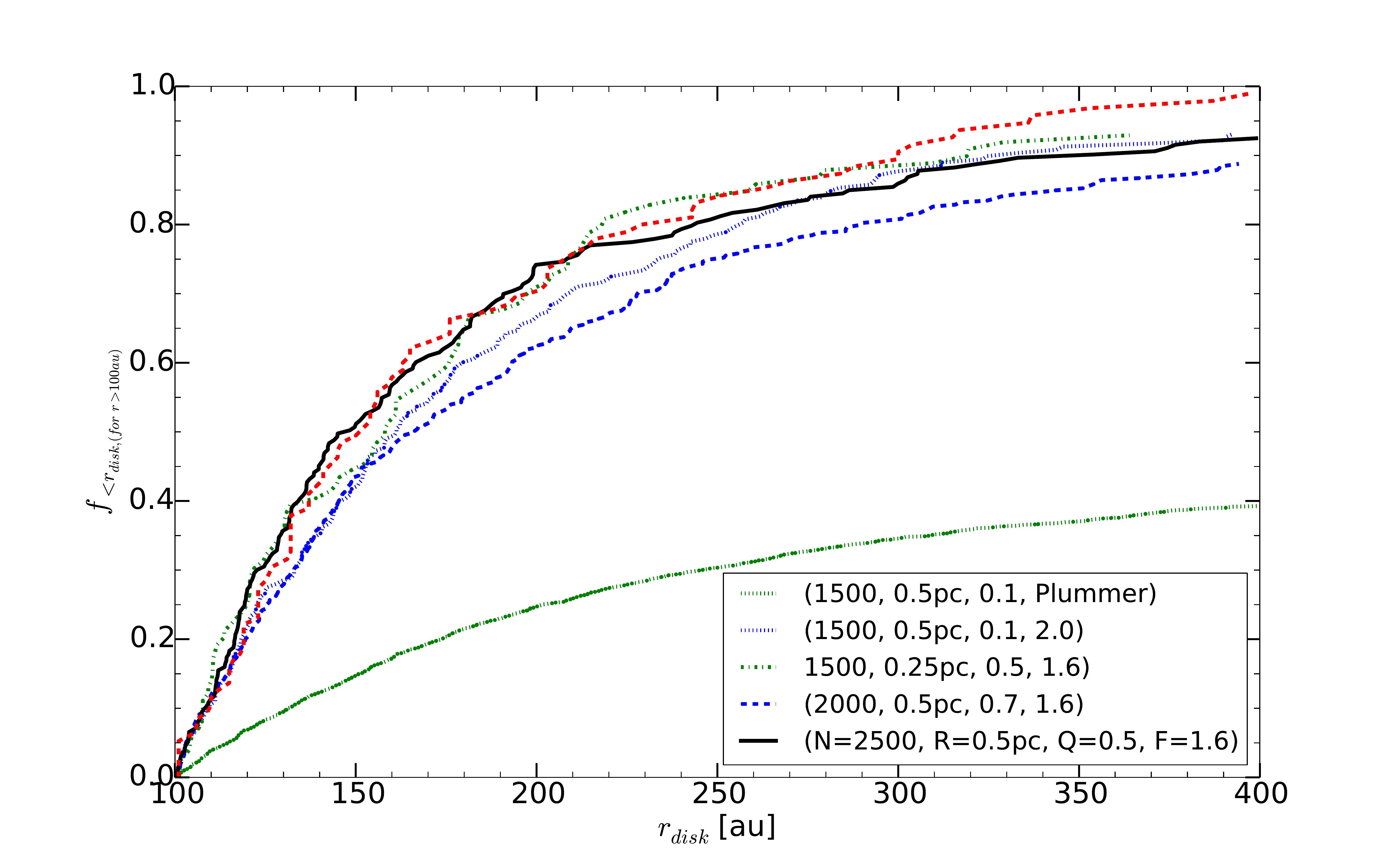}
\end{center}
\caption{Cumulative size distribution of circum stellar disks in the
  Trapezium cluster.  The observations are complete for disk radii
  exceeding 100\,AU. The red dotted curve gives the size distribution
  from the 95 observed disks. The other curves are the result of model
  simulations (with the legend indicating the simulation model
  parameters in the lower right corner).  In contrast to the data
  presented in Tab.\,\ref{Tab:InitialConditions} we only show curves
  at an age of 0.3\,Myr, whereas in the table we present the models
  with the highest KS-$p$ values, which sometimes have a different
  age.
  \label{Fig:ObservedDisDistribution}
}
\end{figure}

In Fig.\,\ref{Fig:ObservedDisDistribution} we present the cumulative
distribution of disk sizes in the simulations and compare them with
the observed distribution. To better compare with the observed
distribution we only present here disk radii of 100\,AU and larger.
The degree of consistency between the observations and simulations is
expressed in the $p$ statistics of the Kolmogorov-Smirnov \citep[][KS
  hereafter]{1954Kolmogorov}, the Mann-Whitney-Wilcoxon \citep[][MW
  hereafter]{1947AMS...18...50M,1945BB...1...80W} tests.
For each of these tests, small values of the statistics $p$ ---say if
$p\aplt 0.05$--- we argue that the two distributions were sampled from
different parent distributions, and these simulations are considered
not to represent the observational data.

In Tab.\,\ref{Tab:InitialConditions} we present the resulting $p$
values for the best simulations. Honestly, it is hard make a
qualitative judgment based on $p$ values alone.  We argue, however,
that the number of disks with a radius $\geq 100$\,AU also have to be
taken into account.  Following Poisson statistics we argue that the
observed number of disks with a radius $\geq 100$\,AU must be between
70 and 120 (if we do not take the statistics of the simulation into
account).  Several of our simulations resulted in satisfactory KS and
MW statistics, but with so few (or so many) stars within the
appropriate range that we could exclude them from further
consideration.  The remaining cases are listed in
Tab.\,\ref{Tab:InitialConditions} (plus three ill comparisons for
completeness, because those are presented also in
figs.\,\ref{Fig:SimulatedDisDistribution} and
\ref{Fig:ObservedDisDistribution}).

\begin{table*}
  \caption{ Results of the simulations. The top 9 rows give the
    initial conditions for the simulations which compare best with the
    observations. The bottom three rows give the results of additional
    simulations, for which the data is also presented in the
    accompanying figures.  The first two column gives the number of
    stars and the time (in Myr) at which the snapshot was compared
    with the observations. The following columns give the initial
    radius of the cluster (in parsec), the virial ratio and the
    fractal dimension. The subsequent three columns give the KS and NW
    $p$ values for the comparison between the observed disk
    distribution and the simulations, and the number of stars with a
    disk radius $\geq 100$\,AU, which corresponds to the observational
    limit The last two columns give the KS $p$ value for the
    comparison between the observed and simulated disk mass, and the
    number of disks in the simulations which comply to the observed
    limits ($m > 4.2$M$_{\rm Jupiter}$, 20\,au $< r <$ 200\,au, see
    \S\,\ref{Sect:DiskMassDistribution}).
  \label{Tab:InitialConditions}
}
  \begin{center}
  \begin{tabular}{llllllllll}
  \hline
     N    &t/Myr & R/pc & Q   & F   & $p_{\rm r, KS}$ & $p_{\rm r, NW}$ & $N_{r \geq 100au}$ & $p_{\rm m, KS}$ & $N_{20<r<200au}$\\
     2000 & 0.1 & 0.25& 0.1 & 2.0 &  0.80 & 0.40 & $ 86 \pm 19$ & 0.30 & $ 236 \pm 17$ \\
     2500 & 0.1 & 0.25& 0.1 & 2.0 &  0.84 & 0.39 & $ 93 \pm 37$ & 0.32 & $ 315 \pm 52$ \\
     2500 & 0.2 & 0.5 & 0.1 & 1.6 &  0.77 & 0.43 & $ 80 \pm 24$ & 0.50 & $ 145 \pm  5$ \\
     1500 & 0.3 & 0.5 & 0.3 & 1.6 &  0.72 & 0.22 & $ 78 \pm 23$ & 0.48 & $ 156 \pm 30$ \\ 
     2500 & 0.4 & 0.5 & 0.3 & 1.6 &  0.68 & 0.34 & $ 74 \pm 27$ & 0.58 & $ 193 \pm 14$ \\
     3000 & 0.3 & 0.5 & 0.3 & 1.6 &  0.70 & 0.50 & $ 67 \pm 6$ & 0.77 & $ 213 \pm 10$ \\ 
     2000 & 0.6 & 0.5 & 0.5 & 1.6 &  0.82 & 0.25 & $ 63 \pm 9$ & 0.76 & $ 155 \pm 15$ \\
     3000 & 0.2 & 0.5 & 0.5 & 1.6 &  0.54 & 0.46 & $ 80 \pm 25$ & 0.63 & $ 223 \pm 30$ \\
     1500 & 1.0 & 0.5 & 1.0 & 1.6 &  0.50 & 0.29 & $ 72 \pm 25$ & 0.46 & $ 144 \pm 11$ \\
\hline
     2000 & 0.2 & 0.5 & 0.7 & 1.6 &  0.02 & 0.01 &  $95 \pm 29$ & 0.30 & $169 \pm 28$ \\
     1500 & 0.3 & 0.25& 0.5 & 1.6 &  0.86 & 0.47 &  $25 \pm 10$ & 0.66 & $ 90 \pm 14$ \\
     1500 & 0.3 & 0.5 & 0.1 & Pl  &  0.00 & 0.00 & $986 \pm 23$ & 0.05 & $353 \pm 17$ \\
\hline
\hline
  \end{tabular}
  \end{center}
\end{table*}

\begin{figure}
\begin{center}
\includegraphics[width=0.5\textwidth]{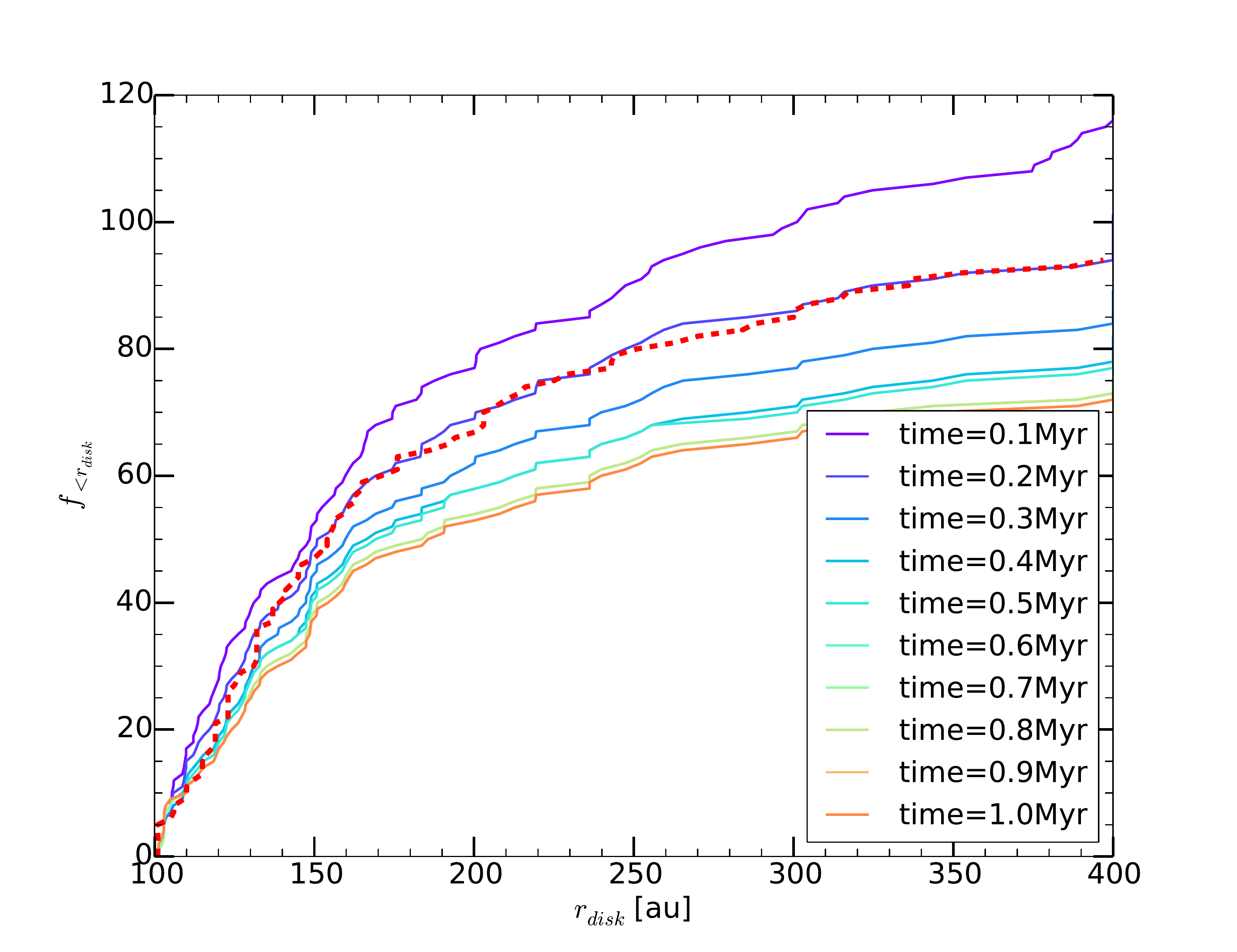}
\end{center}
\caption{Time evolution of the cumulative disk-size distribution for
  one of our simulations with $N=2000$ stars with $R=0.5$\,pc, $Q=0.7$
  and $F=1.6$. This run was not included in
  Tab.\,\ref{Tab:InitialConditions}, because even though this
  particular run gave a very satisfactory comparison with the
  observations, the other 3 runs did not as well and the overall value
  for $p = 0.024$ at $t=0.2$\,Myr.  The distributions are not
  normalized, to show how the number of disk sizes in the observed
  range decreases with time. The normalized version at $t=0.3$\,Myr is
  also presented as the blue dashed curves in
  Fig.\,\ref{Fig:SimulatedDisDistribution} (for the complete
  distribution) and in Fig.\,\ref{Fig:ObservedDisDistribution} for the
  normalized cumulative distribution.
\label{Fig:discevolution}
}
\end{figure}

We demonstrate the evolution of the number of stars with disks $r_{\rm
  disk} \geq 100$\,AU in Fig.\,\ref{Fig:discevolution}, where we
present the time evolution of the distribution of disk sizes for one
of the simulation. At an age of $t=0.2$\,Myr the disk size
distribution compares well with the observed distribution, in shape as
well as in number. At later age the number of stars with disks in the
appropriate regime drops quite dramatically.

\begin{figure}
\begin{center}
\includegraphics[width=0.5\textwidth]{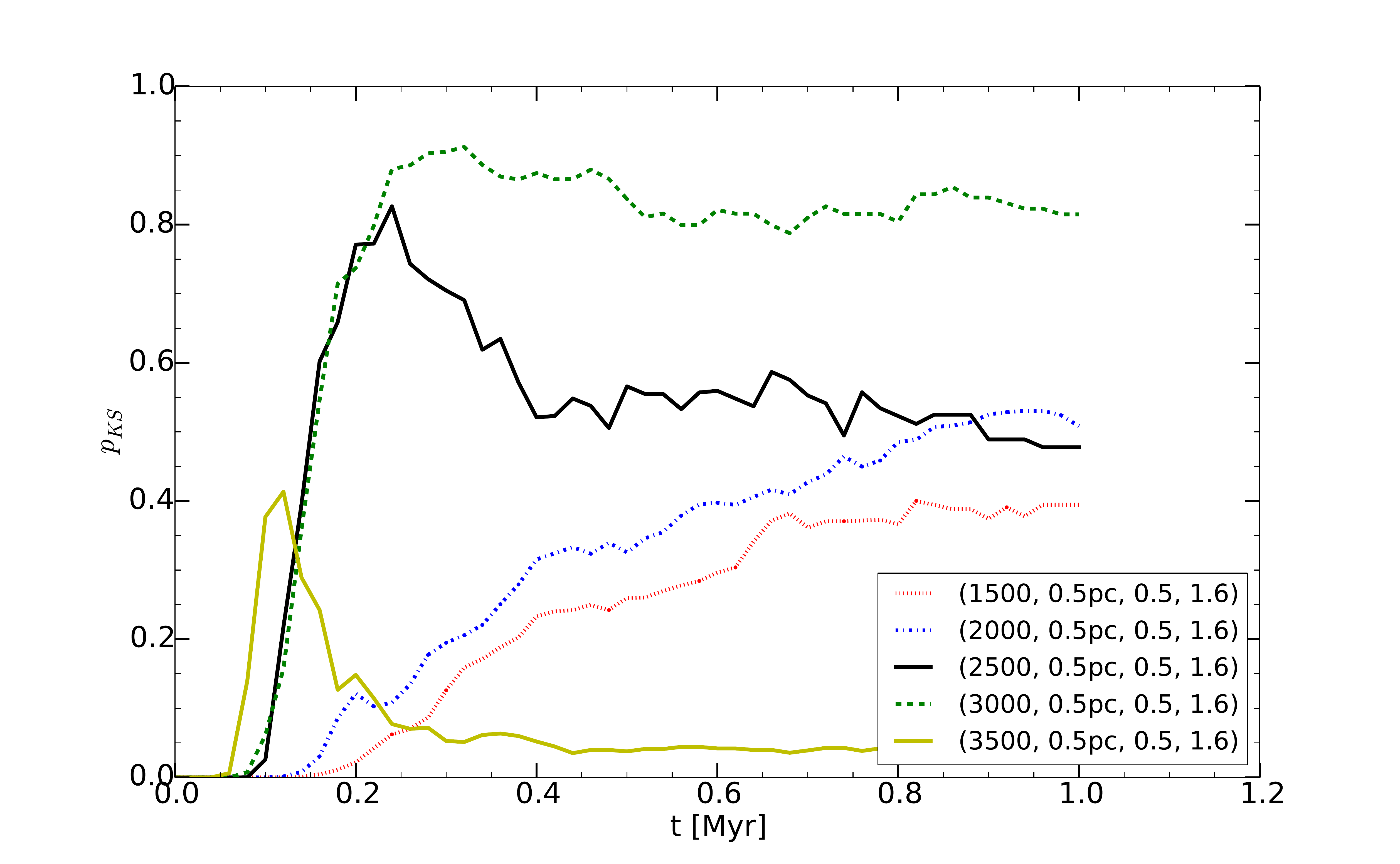}
\end{center}
\caption{Kolmogoriv-Smirnov $p$ value for simulation with $N=1500$ to
  $N=3500$ stars with $R=0.5$\,pc, $Q=0.5$ and fractal dimension
  $F=1.6$ up to an age of 1\,Myr.  The KS values are calculated by
  summing over 12 runs for each set of parameters with a time
  resolution of 0.02\,Myr.
\label{Fig:KSstat_for_R05Q05F16.pdf}
}
\end{figure}

In Fig.\,\ref{Fig:KSstat_for_R05Q05F16.pdf} we present the time
evolution of several simulations with a characteristic radius of
$R=0.5$\,pc, a virial temperature of $Q=0.5$ and fractal dimension
$F=1.6$. Those simulations generally result in the highest $p$ values
for KS statistics (and equivalently so for for MW statistic) in
comparison with the observed circum stellar disk-size distribution.

The runs with $N = 1000$ and 2000 show a steady but relatively slow
rise to a KS probability $p\simeq 0.5$ on a time scale of about
1\,Myr, and the high $N=3500$ simulation peaks at $t\simeq 0.1$\,Myr
but hardly exceeds $p=0.4$.  The simulations with intermediate $N$
(between 2500 and $N=3000$) show a promising trend of peaking around
$t=0.2$---0.3 with a maximum $p \sim 0.9$.

\subsection{Disk mass distribution}\label{Sect:DiskMassDistribution}

Disk masses have been determined in the Orion Trapezium cluster using
millimeter observations \citep{2009ApJ...694L..36M}.  This has been
calculated from the spectral energy distribution from centimeter to
submillimeter wavelengths and of the interferometric response to the
cloud background for 26 out of 55 HST-identified proplyds
\citep{2009ApJ...694L..36M}.  They show that the number of disks per
logarithmic mass interval is approximately constant over almost a
decade in mass between 4.2 and $35.6$\,M$_{\rm Jupiter}$.  Because
these disks were selected to have bright millimeter emission, the
sample is biased toward relatively large disks between 20 and 200\,AU
\citep{2009ApJ...694L..36M}.  Our rather limited understanding of the
initial disk mass, the radial density profile and therefore of the
effect of encounters on these disks, limits the validity of comparing
the observations with the simulations.

\begin{figure}
\begin{center}
\includegraphics[width=0.5\textwidth]{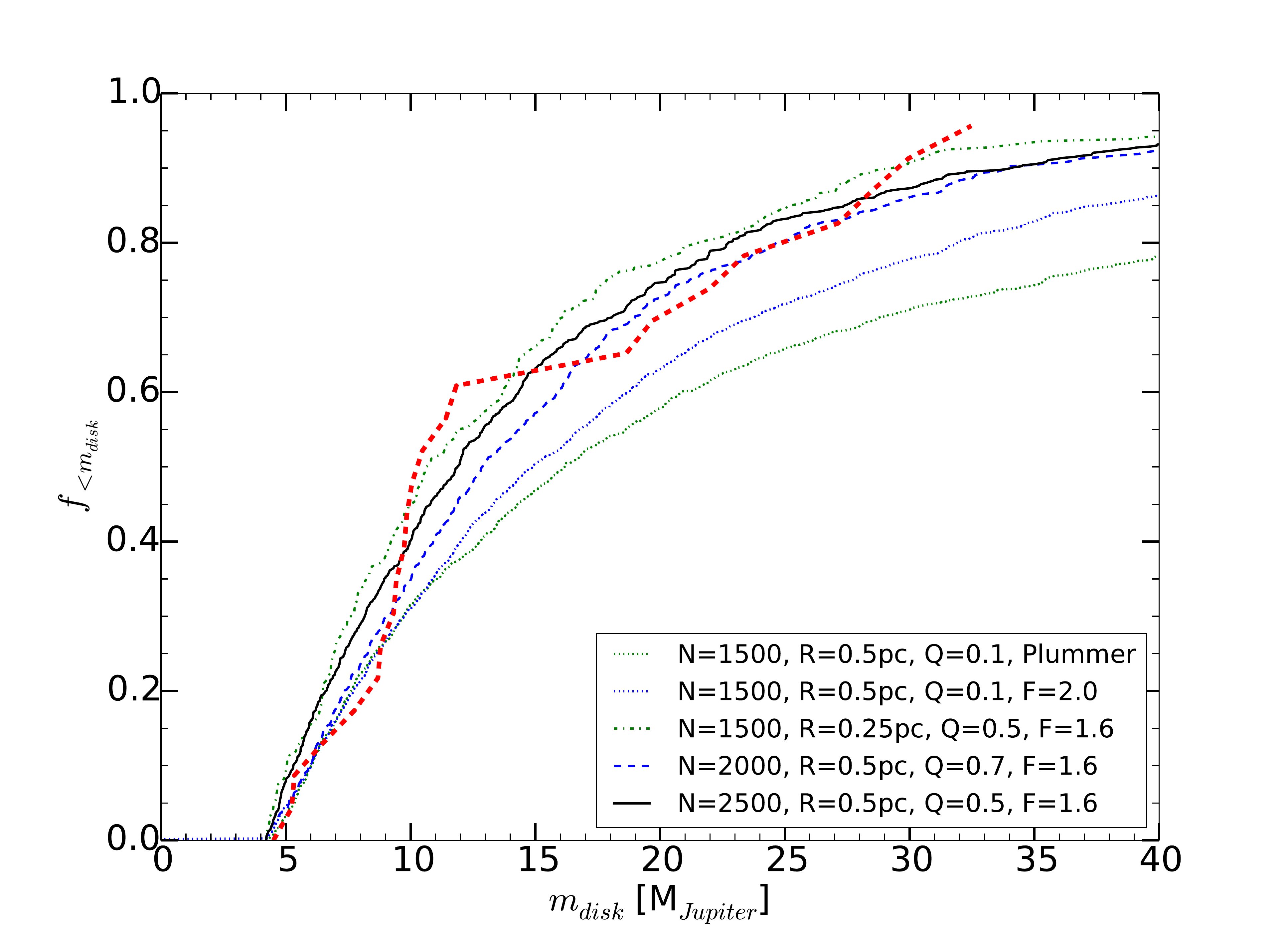}
\end{center}
\caption{Cumulative distribution of disk masses in the Trapezium
  cluster (using $f=1$ in Eq.\,\ref{Eq:dmdisk}).  The observations are
  claimed to be complete down to a mass of 4.2\,M$_{\rm Jupiter}$,
  below which we do not plot any data.  There are only 23 observed
  disks with at least this mass.  The other curves give the result of
  the model simulations at an age of 0.3\,Myr; the legend (bottom
  right) explains the initial model parameters, but they are the same
  as in Fig.\,\ref{Fig:SimulatedDisDistribution} and
  \ref{Fig:ObservedDisDistribution}. Except the two dotted curves
  (green and and blue), each of these models produce a satisfactory
  fit to the observed disk masses.  Due to the limited statistics, the
  disk mass distribution generally compares much better to the
  observations than the disk size distribution.
\label{Fig:SimulatedMassDisDistribution}
}
\end{figure}

In Fig.\,\ref{Fig:SimulatedMassDisDistribution} we present the
cumulative mass distribution of several simulations and compare them
with the observed mass distribution. Apart from the Plummer initial
conditions, it appears to be difficult to exclude any of the model
simulation in Tab.\,\ref{Tab:InitialConditions}.  Our assumption of an
initial disk mass of 10\% of the zero-age stellar mass is quite
arbitrary. If we would have adopted half this value, the curves in
Fig.\,\ref{Fig:SimulatedMassDisDistribution} skew somehwat to the left
the tree overlapping curves (black, blue dashed ans green dash-dotted
cuves) just staying above the observed distribution, whereas the blue
dotted curve (N=1500, R=0.5pc, Q=0.1, F=2.0) compares best with the
observations.

\begin{figure}
\begin{center}
\includegraphics[width=0.5\textwidth]{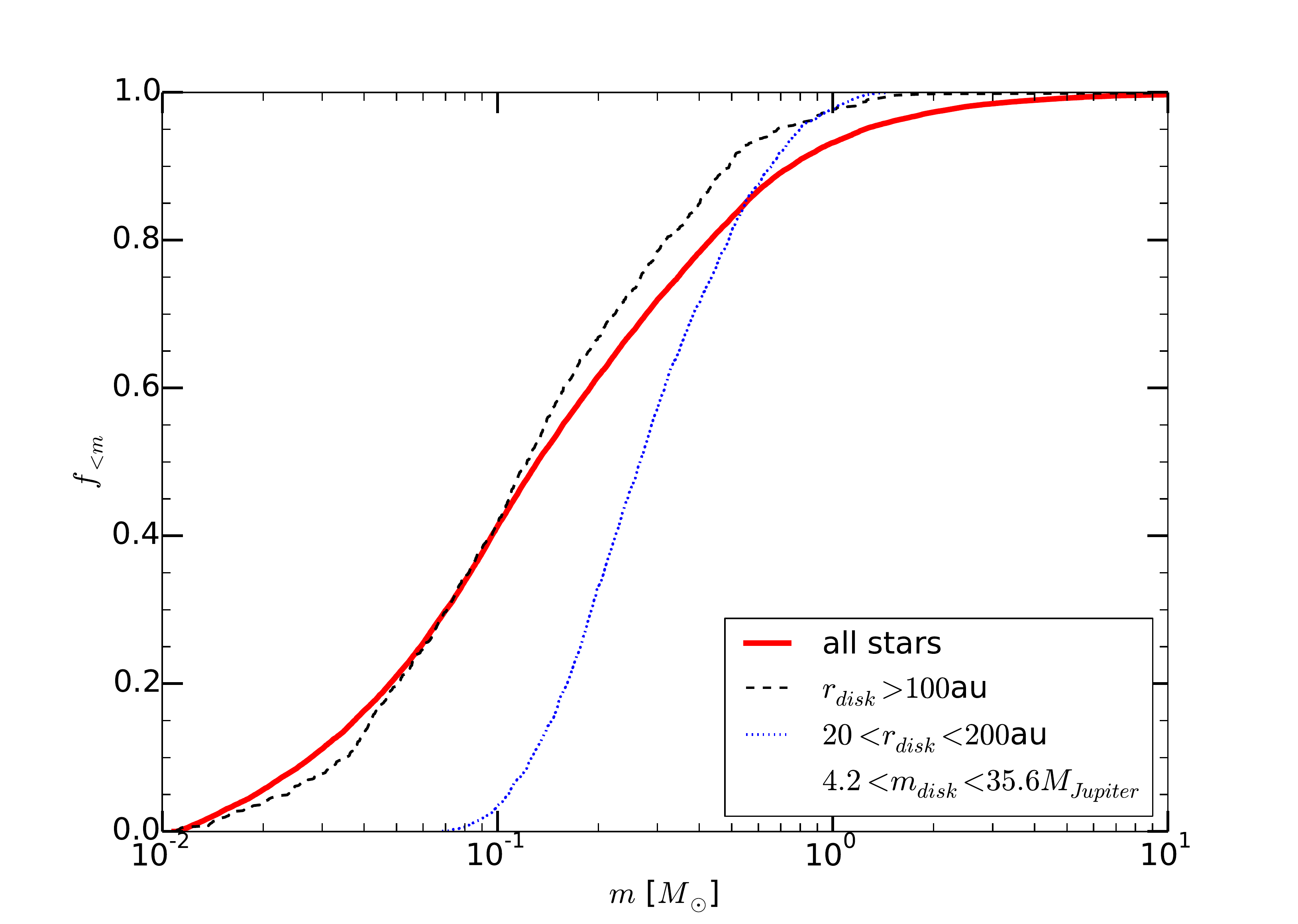}
\end{center}
\caption{Cumulative distribution of stellar masses in simulation with
  $N=2000$ stars, $R=0.5$\,pc, $Q=0.5$\,pc and $F=1.6$ at an age of
  0.3\,Myr.  The red solid curve gives the mass function for all
  stars, and is identical to the initial mass function adopted for all
  our simulations. The black dashed curve give the mass function for
  the stars with a disk $r_{\rm disk}>100$\,AU, and the blue dotted
  curve for the stars with a disk size between 20 and 200\,AU and with
  a disk mass between $4.2\,M_{\rm Jupiter}$ and $35.6\,M_{\rm
    Jupiter}$.
\label{Fig:SimulatedStellarMassfunction}
}
\end{figure}

In Fig.\,\ref{Fig:SimulatedStellarMassfunction} we compare the mass
functions for the all stars with those with a selected disk sizes and
masses. Stars with a massive disk ($m_{\rm disk} > 4.2$\,M$_{\rm
  Jupiter}$) have considerably higher massed ($\langle m \rangle
\simeq 0.34$\,\MSun) whereas those with a relatively large disk
($r_{\rm disk} > 100$\,au) tend to be somewhat less massive than
average ($\langle m \rangle \simeq 0.22$\,\MSun).

The majority of simulations with rather cold initial temperature
($Q=0.1$) fail to reproduce the observed disk distribution. Exceptions
are simulations born with a fractal dimension of 1.6, which is also
preferred for the disk size distribution. Most simulations with a
virial temperature of $Q=0.3$ and $Q=0.5$ provide a satisfactory
comparison with the observations.

\section{Discussion and conclusions}

We have performed simulations of self gravitating stellar systems in
which we incorporated a semi-analytic prescription for the effect of
encounters on the sizes and masses of circum-stellar disks.  Our
simulations aim at reproducing the disk size distribution observed
around 95 stars in the Trapezium cluster \citep{2005A&A...441..195V}
and with the 23 observed disk masses \citep{2009ApJ...694L..36M}.

Our simulations ignore most physical effects that tend to play an
important role in the evolution of these systems, these include the
presence of residual gas from the parent molecular cloud, the tidal
field of the Galaxy, primordial binaries and mass segregation, stellar
evolution and feedback process.  The only aspect we take into account
properly in our simulations are the gravitational encounters between
stars. We estimate, at run time, how circum stellar disks are
truncated as a result of two-body encounters.  Regardless of the
limited physics in our simulations the resulting disk-size
distribution compares excellently with the observations. The disk mass
distribution gives a consistent picture, but is less constraining due
to the smaller number of observed masses.  Bolstered by our success to
reproduce the observed disk radius distribution from the simple
argument that this distribution originates from close encounters in
the young star clusters, we use this argument to limit some of the
structure parameters of the cluster at an earlier age.

The disk-size distribution appears to match the observations best if
the Trapezium cluster was born virialized ($Q=0.5$) or slightly colder
($Q=0.3$) and with a half-mass (or characteristic) radius of $R \simeq
0.5$\,pc.  Due to the large run-to-run variations it is hard to
constrain these values further. With these parameters the best
comparison is achieved for clusters with $N=2500$ to 3000 stars.

Simulations with a smaller number of stars tend to under produce the
number of stars in the range where disks have been observed, whereas
in more massive cluster disks tend to be harassed on too short a time
scale.

We exclude a Plummer sphere as the initial density profile,
irrespective of the initial cluster radius, because too few disks are
truncated. Even with cold initial conditions $Q=0.1$ and a tiny radius
$R=0.125$ the Plummer sphere will always have some stars that remain
insufficiently affected by dynamical encounters within the available
time.  For similar reasons we also exclude the density distribution
with a relatively high fractal dimension $F\apgt 2$, and cluster with
a large characteristic radius of $R\apgt 0.7$\,pc, irrespective of the
kinematic temperature of the initial cluster.  Clusters with a fractal
initial distribution of stars, with relatively cold initial conditions
or a small characteristic radius $R\aplt 0.3$\,pc are also excluded,
because they truncate too many disks too effectively.

An initial density distribution generated using a fractal dimension of
$F=1.6$ result in the best comparison with the observations, although
the number of stars should then be between $N=2500$ and 3000, the
virial temperature of $Q\simeq 0.3$---0.5, and with a half mass radius
of $R\simeq 0.5$\,pc.

These results are robust against small changes in the initial
conditions.  Additional simulations with a Salpeter mass function with
a lower mass limit of 0.1\,\MSun, have no appreciable effect on the
results.  Our prescription for the mass evolution in close encounterd
depends on the initially adopted disk size, for which we adopted
400\,AU. We performed additional simulations with initial disk radii
of 1000\,AU, but this had no appreciable effect on the resulting disk
mass size distribution.

We have not studied the effect of primordially
mass-segregation, but we think that the effect somewhat mimics the a
slight reduction in the initial kinematic temperature. It could
therefore be preferrable to start with slightly warmer initial
distributions if the degree of initial mass segregation is
appreciable.

The distributions of disk sizes and masses for stars that have escaped
the cluster are not appreciably different than those that remain
bound, irrespective of the age of the cluster. Stars tend to escape
after a strong encounter and the disks have already been truncated by
that time.

\subsection{Further considerations}

The Solar system may have been truncated at about 35\,AU by a close
encounter with another star \citep{2015MNRAS.453.3157J}.  According to
our calculations a truncation between 10\,AU and 100\,AU occurs in
about 25\% of the planetary systems born in a cluster with parameters
similar to the Trapezium cluster.  The parameters of the Trapezium
clusters, as constrained by our calculations, are not inconsistent
with the possible parameters of the cluster in which the Sun was born
\citep{2009ApJ...696L..13P}, although there the anticipated cluster
was slightly larger, $2\pm1$\,pc and probably somewhat more massive
($2000 \pm 1000$\,\MSun). We still favor this larger cluster radius
for forming the Solar System, because of the need to also survive the
ablation of the protoplanetary disk by the ablation of supernovae.
This process happens at a somewhat later time ($\apgt 5$\,Myr), and is
not accounted for in our simulations.

With the parameters that give a best comparison with the Trapezium
cluster, $\sim 70$\% of disks are truncated below $\sim 10$\,AU within
the first few hundred thousand years of their dynamical evolution.
Further truncation may be initiated by photoevaporation of the massive
stars in the young cluster \cite{2004ApJ...611..360A}, but that does
not effect the earlies evolution studied here. Violent disk truncation
may not particularly hinder the planet formation process, but it sure
excludes the formation of planets in obits beyond the disk truncation
radius. Another aspect of the truncation of protoplanetary disks may
be the exclusion of Earth-like ice giants, which are expected to form
well beyond the ice line ($a\apgt 10$\,AU), and subsequently sink
closer to the star in the remaining disk via what is called the grand
tack model \citep{2014Icar..232...81M,2015DDA....4630004D}

In a recent study \cite{2015A&A...577A.115V} conclude that mutual
stellar encounters are responsible for truncating proto planetary
disks in young small clusters; in clusters with an average density of
60\,star/pc$^3$, such as the Orion nebula cluster, up to 65\% of
protoplanetary disks are truncated below 1000\,AU, whereas 15\% is
truncated even below 100\,AU.  In a more dense environment
(500\,star/pc$^3$) these fractions increase to 85\% and 39\%.  Our
calculations are consistent with the analysis of
\cite{2015A&A...577A.115V}, but in order to reproduce the disk-size
distribution observed in the Orion Trapezium cluster the initial
cluster density has to be much higher, $\sim 10^3$\,star/pc$^3$.

It is interesting to note that all stars in the simulations that
reproduced the disk-size distribution observed in the Orion Trapezium
cluster have captured some material from the disks of other stars. In
our most favorite simulation for the Trapezium cluster ($N=2000$,
$R=0.5$\,pc, $Q=0.5$, $F=1.6$), about 60\,\% of the stars have
captured $\apgt 1$\% of their own disk mass from another star in a
close encounter, and $\sim 34$\% captured more than 10\% of mass.
This is consistent with the idea proposed by
\cite{2015MNRAS.453.3157J} for the origin of the planetesimal Sedna,
being captured from another star in the early evolution of the Solar
System.

\section*{Acknowledgments}

We thank Anthony Brown, Michiel Hogerheijde, Lucie J\'{i}lkov\'{a} and
Igas Snellen for discussions. This work was supported by the
Netherlands Research Council NWO (grants \#643.200.503, \#639.073.803
and \#614.061.608) by the Netherlands Research School for Astronomy
(NOVA). The numerical computations were carried out on the Little
Green Machine at Leiden University.




\end{document}